\documentclass[11pt]{article}
\pdfoutput=1 

\usepackage[T1]{fontenc}
\usepackage[utf8]{inputenc}
\usepackage{lmodern}

\usepackage[margin=1in]{geometry}
\usepackage{setspace}
\usepackage{amsmath,amssymb,amsthm}

\usepackage{booktabs}
\usepackage{tabularx}
\usepackage{array}

\usepackage{graphicx}
\graphicspath{{./}}
\usepackage{subcaption}
\usepackage{caption}
\usepackage{listings}
\usepackage{microtype}

\usepackage[colorlinks=true,allcolors=blue]{hyperref}

\usepackage{tikz}
\usetikzlibrary{positioning, arrows.meta, shapes.geometric, fit, backgrounds}

\title{Taming Scylla: \\
Understanding the multi-headed agentic daemon of the coding seas}
\author{Micah Villmow\thanks{This paper combines the author's original research and writing with improvements, suggestions, and rewrites provided by Claude Code (claude.ai/code).} \\
Individual \\
research@villmow.us}
\date{}

\begin{document}

\maketitle

\begin{abstract}
LLM-based tools are automating more software development tasks at a rapid pace, but
there's no rigorous way to evaluate how different architectural choices—prompts,
skills, tools, multi-agent setups—materially affect both capability and cost.
This paper introduces Scylla, an evaluation framework for benchmarking agentic
coding tools through structured ablation studies that uses seven
testing tiers (T0-T6) progressively adding complexity to isolate
what directly influences results and how.
The key metric is Cost-of-Pass (CoP): the expected dollar cost to get one
correct solution, which directly quantifies the trade-off between complexity and
efficiency.
The framework is model-agnostic, designed to work with any CLI tool; this paper
demonstrates it with Claude Sonnet 4.5, using multiple LLM judges (Opus 4.5,
Sonnet 4.5, Haiku 4.5) from the same vendor for evaluation consensus, where judges
score results using direct tests, human-designed LLM-evaluated rubrics, and qualitative
assessment.
The result is a reproducible framework that quantifies trade-offs between agent
complexity and actual outcomes, suggesting that architectural complexity does not always improve quality.
\end{abstract}

\noindent\textbf{Keywords:} LLM agents, software engineering benchmarks, cost-of-pass, multi-agent systems,
prompt engineering, ablation studies, evaluation frameworks, CLI tools, agentic
AI

\section{Introduction}\label{sec:intro}

Large language models have ushered in massive increases in capabilities for
automated computer interactions. What used to require hand-coded algorithms and
pipelines can now be done automatically using state-of-the-art coding models.
However, understanding what improves these language models is more black
magic than art, let alone a rigorous science. This paper's goal is to help
demystify the magic of prompt engineering by proposing a rigorous evaluation
framework to quantify the benefits of different architectural approaches.

There are benchmarks for measuring LLM workflows in various domains, such as
agent-bench\cite{liu2023agentbench}, swe-bench\cite{jimenez2024swebench}, and tau-bench\cite{yao2024taubench}. There are also prompt
evaluation benchmarks such as PromptBench\cite{zhu2024promptbench} and PromptEval\cite{polo2024efficient}. This paper
focuses specifically on coding tools, particularly the industry-leading Claude
Code\cite{anthropic2024claude}, and how prompt and architectural modifications affect model behavior.
This paper introduces Scylla, a framework for evaluating agentic coding tools in
a systematic way, allowing extension to domains beyond CLI-based coding tools.
The dryrun results on a trivial Hello World task show that all seven tiers
(T0-T6) achieve equivalent quality (all grade A, scores 0.943-0.983) while cost
varies 3.8$\times$ from \$0.065 (T5 hybrid) to \$0.247 (T6 super). The framework
successfully differentiates cost structures across architectural choices even
when quality converges, demonstrating that architectural complexity does not
always improve quality. Careful hybrid designs (T5) can achieve Frontier
Cost-of-Pass (the minimum CoP across all tiers) by selectively combining features rather than maximizing them.

Anthropic has many good resources for improving Claude Code on their engineering
blog, but despite these, there are not any intuitive and user-friendly methods
for comparing whether changes to the prompt instructions will yield tangible
benefits. Therefore, I am introducing Scylla, a testing framework for evaluating
prompts, tools, skills, and agents for solving problems that are common for
day-to-day coding tasks. I wanted to know if sub-agents, skills, tools, or MCP
servers were contributing to actual improved code output, without relying on my
gut or intuition. This problem came up multiple times when asked by others to
explain how to better utilize CLI tools for programming. In my experience, the
quality of the prompts has a dramatic effect on the output.
Whether it is the prompt to call the tool or MCP server, the prompt to spawn a
sub-agent, or the prompt to trigger a skill, these language-based triggers are
fuzzy in their meaning. Unlike a traditional programming language that is very
explicit in what it means and what it does, prompts do not map directly and
consistently to action. This framework is my attempt at helping unwrap this
problem.

The remainder of this paper is organized as follows. First, in Section~\ref{sec:related}, I will introduce the current work that is being done in
this area, and explain how they approach the problem. Then, in Section~\ref{sec:methodology}, I will
introduce the testing methodology along with an in-depth analysis of the first
test case. This will provide the needed understanding of what is being tested,
along with why, on something that is easily analyzed and
understandable. Section~\ref{sec:architecture} describes the framework's software architecture, explaining how the evaluation system is built and how it achieves model-agnostic evaluation through the adapter pattern. Then I will go over the rest of the testing framework to
showcase what is being tested, measured, and why these are being tested using
simple cases introduced in the previous sections.

The framework is designed to investigate questions such as these in future studies:

\begin{itemize}
\item Whether it is possible to quantify whether a task is solvable more efficiently by one methodology over others
\item Whether the sum of a prompt is more than the individual parts
\item Whether there are core improvements that can be made purely through extensions to Claude Code that are generic for all workloads
\item Whether there are specific prompt techniques that have secondary effects, positive or negative, on other prompt techniques
\item Holding the tool and prompt constant, how much the underlying model may contribute to the quality of the results
\end{itemize}

Some hypotheses the framework is designed to test in subsequent work include:

\begin{itemize}
\item \textbf{(H1)} Certain tasks may excel when run as sub-tasks, tools, MCP servers, or skills, for reasons unrelated to context management
\item \textbf{(H2)} Prompt complexity may have opposing effects on quality: KISS principle (negative correlation) for scenarios in the training set versus inverse KISS (positive correlation) for scenarios outside the training set
\end{itemize}

\section{Related Work}\label{sec:related}

Given that we are testing production tools and not models, many, if not all, of
the prior work on evaluating prompts and benchmarks does not apply here. Since
there is possibly a large level of indirection between what we are testing and
what actually gets executed by the model due to engineering trade-offs, I am
considering the tool to be a black box and not attempting to reverse engineer
this tool. Despite this, what is executing is hidden behind multiple layers,
first being the CLI tool itself, but also whatever optimizations and
implementation details the vendor implements on top of their trained base model.
The models themselves are not fully documented publicly, as these details are
competitive advantages, and the pre or post-processing that occurs is not always
visible to the user as they can occur vendor-side.

There are several good benchmarks for evaluating LLM agents and prompts. SWE-Bench\cite{jimenez2024swebench} tests
models on real GitHub issues. Agent-Bench\cite{liu2023agentbench} tests multi-turn agents across
different environments like operating systems, databases, and knowledge graphs,
with fine-grained metrics beyond pass/fail. TAU-Bench\cite{yao2024taubench} focuses on whether
agents can effectively use external tools. For prompt evaluation, there is PromptBench\cite{zhu2024promptbench} (unified testing across tasks),
PromptEval\cite{polo2024efficient} (automated correctness and robustness checking), and EleutherAI's
lm-evaluation-harness\cite{gao2024lmevalharness} (standardized multi-task comparison).

However, these benchmarks all assume direct access to model inputs and outputs, evaluating models
directly rather than complete CLI tools. With production CLI tools like Claude Code, the model is wrapped in
layers of system prompts, tool schemas, skill definitions, hooks, skills, MCP servers,
vendor optimizations, and orchestration logic. I cannot just test the model in isolation, so I must test the whole
system. No standard benchmarks exist for CLI tools like Claude Code and how prompts affect them.

My work is based solely on evaluating CLI tools, as the CLI's tools are more
than the model themselves. As I mentioned earlier, the agentic loop, with hooks,
tools, skills, sub-agents, MCP servers, and other logic wrapped together into a
single application where the only way to get control of the behavior is through
the English language is what I want to evaluate for effectiveness. From this
interface, programmatic tools can be spawned, but the ability to properly and
accurately interact with the agent is via a fuzzy language interface, and not
via traditional programmatic interfaces. While there are some hooks that allow
extra programmatic validation with Claude Code, I am not evaluating those at this
time. Claude Code has the ability to use agentic evaluation at the hook
boundary, but triggering it is guaranteed (and not language-based), so it is not
interesting for probabilistic evaluation.

\section{Test Methodology}\label{sec:methodology}

\subsection{Experimental Design}

This experiment is designed by testing English phrases, colloquially known as
prompts, via the various methodologies exposed by a CLI tool, in this case
Claude Code. The prompts to be tested are taken from the ProjectOdyssey\cite{projectodyssey} git
repository at GitHub hash 011a3ff on December 30th, 2025. The prompts are broken
down into their components and separated into various tiers which will be
discussed later. These components are used to set up the experiment, which is run
by allowing an agent a nearly unfettered access to the system, only blocking
dangerous ops, thanks to the safety-net plugin\cite{safetynet} from cc-marketplace\cite{ccmarketplace}, to
perform a task. The task has a well defined solution that is then judged by
three different LLMs of various 'strength'. In this case Claude Opus 4.5,
Claude Sonnet 4.5, and Claude Haiku 4.5. Each of the 4.5 models are sufficiently
advanced in capabilities to be considered independent judges of a task with low
failure rates. The judges are provided the same prompt, so the only difference
between their results comes from the judge training and implementation
differences and not from the prompt or test input. Each judge will receive the
output of the task LLM, and provide the results based on the criteria. The
judges have the following categories of evaluation; functional correctness, code
quality, proportionality, build pipeline, and overall quality.

\begin{table}[htbp]
\centering
\caption{LLM-as-Judge Evaluation Categories}
\begin{tabularx}{\textwidth}{llX}
\toprule
Category & Weight & Description \\
\midrule
Functional Correctness & 0.35 & File existence, output correctness, exit codes, exact output matching \\
Code Quality & 0.20 & Syntax validity, idiomatic code, unused imports, PEP8 compliance  \\
Proportionality & 0.15 & Appropriate scope, minimal files, no unnecessary artifacts or tests  \\
Build Pipeline & 0.10 & Build passes, format checks, tests (when applicable), pre-commit hooks  \\
Overall Quality & 0.20 & Engineering judgment on appropriateness, maintainability, and senior engineer approval \\
\bottomrule
\end{tabularx}
\end{table}

\textbf{Total Weight}: 1.0 (100\%)

Each category contributes proportionally to the final score. Here is the formula:

$$S_{final} = \sum_{i} w_i \cdot \frac{P_i^{achieved}}{P_i^{max}}$$

where $w_i$ are the category weights (they sum to 1.0), and $P_i$ is the points
the test got versus the maximum possible (skipping any N/A items). For scoring
individual items:

\begin{itemize}
\item \textbf{Binary items}: You either get it or you do not (1.0 or 0.0)
\item \textbf{Graduated items}: Partial credit on a 0.0-1.0 scale based on results
\item \textbf{Subjective items}: LLM judgment with calibrated deductions

\end{itemize}

\begin{table}[htbp]
\centering
\caption{Deduction Calibration Scale (for subjective assessment)}
\begin{tabular}{lll}
\toprule
Severity & Deduction Range & Examples \\
\midrule
Negligible & 0.00-0.05 & IDE config files, \texttt{\_\_pycache\_\_} artifacts \\
Trivial & 0.05-0.15 & Missing trailing newlines, unused imports \\
Minor & 0.15-0.30 & Missing docstrings, magic numbers \\
Moderate & 0.30-0.50 & Code duplication, hardcoded values \\
Major & 0.50-0.80 & Non-critical security issues, race conditions \\
Severe & 0.80-1.50 & Critical security vulnerabilities \\
Critical & 1.50+ & Non-functioning solutions, destructive operations \\
\bottomrule
\end{tabular}
\end{table}

The final score maps to a grade using this scale:

\begin{table}[htbp]
\centering
\caption{Grade Scale}
\begin{tabular}{llll}
\toprule
Grade & Threshold & Label & What It Means \\
\midrule
S & 1.00 & Amazing & Perfect score, goes above and beyond \\
A & $\geq$ 0.80 & Excellent & Production ready \\
B & $\geq$ 0.60 & Good & Works well, minor tweaks needed \\
C & $\geq$ 0.40 & Acceptable & It works but has issues \\
D & $\geq$ 0.20 & Marginal & Lots of problems but salvageable with effort \\
F & < 0.20 & Failing & Complete failure of task \\
\bottomrule
\end{tabular}
\end{table}

I use \textbf{0.60} (Grade B) as the pass threshold. That means the solution works
and meets requirements, even if there is room for minor improvements. An S grade
needs a perfect 1.00 and you have to actually exceed what was asked for. I would
not expect many, if any, tests to get an S rating.

Each experiment can be reproduced by running the top-level test run script,
which will launch the same set of tasks with the same parameters, where the only
variation is the judgement of the LLM judges when determining how to judge the
work.

This finishes the summary of a single test. However, the tests themselves are
defined differently. The tests are a prompt and a configuration file that specify
a repository, a GitHub hash, a set of configuration files to override any
pre-defined tooling, set of commands to validate the results, and a git worktree to
run everything in to help with reproducibility. The first test is being used as
an example in this paper, and also as a pipe-cleaner to show that everything
works as expected. This example is 'hello world' from octocat, but forked to my
repository just to make sure that the repository is not polluted. The precaution
is done just in case the agents make mistakes or do things that the original
author probably does not want to be bothered by.

\subsubsection{Test-001: Hello World Baseline}

First, let us look at the simplest possible test to make sure everything works.
This is literally just creating a "Hello World" script, which is a pipe-cleaner
for the infrastructure and to discuss the methodology without intermixing with
the complexity of more realistic tests.

\begin{center}
\textbf{Test Configuration:}

\begin{tabular}{ll}
\toprule
Field & Value \\
\midrule
ID & \texttt{test-001} \\
Name & Hello World Task \\
Timeout & 300 seconds \\
Pass Threshold & 0.60 (Grade B) \\
\bottomrule
\end{tabular}
\end{center}

\textbf{Task Prompt:}

Create a Python script \texttt{hello.py} that prints "Hello, World!" to stdout, exits
with code 0, and uses relative paths. The script is created in the
current working directory.

\textbf{Expected Output:}

\begin{lstlisting}
Hello, World!
\end{lstlisting}

\textbf{Expected Result:}

\begin{lstlisting}
print("Hello, World!")
\end{lstlisting}

or

\begin{lstlisting}
#!/usr/bin/python3

print("Hello, World!")
\end{lstlisting}

\begin{table}[htbp]
\centering
\caption{Rubric Categories and Weights}
\begin{tabularx}{\textwidth}{llX}
\toprule
Category & Weight & Key Criteria \\
\midrule
Functional Correctness & 35\% & File \texttt{hello.py} exists; running \texttt{python hello.py} prints "Hello, World!" \\
Code Quality & 20\% & Valid Python syntax; idiomatic code; no unused imports; appropriate structure \\
Proportionality & 15\% & Total files $\leq$ 3; LOC $\leq$ 3; no unnecessary test files; appropriate scope \\
Build Pipeline & 10\% & Syntax check passes; format check passes (if ruff present); no linter errors \\
Overall Quality & 20\% & Senior engineer approval; appropriately scoped for task complexity \\
\bottomrule
\end{tabularx}
\end{table}

\textbf{What Should Happen:}

Even T0 (no system prompt at all) is expected to get an 'A',
since we are talking $\geq$ 0.80 scores. If T0 cannot do Hello World, I will assume
that something is fundamentally wrong with the framework itself and throw out
the results. Higher tiers (T1-T6) are also expected to succeed, as there is no reason
fancy prompts or multi-agent setups would help with something this simple.
However, if performance drops on this test, it means the added complexity is
actually making things worse even on something so simple, so if this happens, we
will analyze why.

Now that we have gone over the test itself, let us discuss the strategy and tiered
approach. The first thing to test is with no prompt at all, including no system
prompt, if the tool allows it. This is to provide as close to a baseline as the
base model as possible by overwriting the system prompt with an empty string and
not using any configuration or non-default settings from the tool. This provides
the baseline that all improvements are measured against. For something as simple
as hello world, this baseline will solve the task. The test setup is such that
variability in judging will occur, but there is not much one can do to improve
the output of a hello world script. However, there are things that you can do
that make things worse or break the expected behavior, but I would expect all
solutions to be the exact same for all the tests. Divergence points to
interesting results.

\subsubsection{Tiered Ablation Strategy}\label{sec:tiered-ablation}

The core idea is simple: start with nothing, then add one set of things at a
time to see what actually helps. This ablation study uses seven tiers that
progressively add complexity, with \textbf{113 subtests} total. Each tier gets
tested independently so we can isolate what each component contributes.

\begin{table}[htbp]
\centering
\caption{Testing Tiers (Ablation Study Framework)}
\begin{tabularx}{\textwidth}{llXllX}
\toprule
Tier & Name & Subtests & Primary Focus & Tools & Delegation \\
\midrule
T0 & Prompts & 24 & System prompt ablation (empty $\rightarrow$ full) & - & No \\
T1 & Skills & 10 & Domain expertise via installed skills & Default & No \\
T2 & Tooling & 15 & External tools and MCP servers & Yes & No \\
T3 & Delegation & 41 & Flat multi-agent with specialists & Yes & Yes \\
T4 & Hierarchy & 7 & Nested orchestration with orchestrators & Yes & Yes \\
T5 & Hybrid & 15 & Optimal combinations from all tiers & Yes & Yes \\
T6 & Super & 1 & Maximum capability configuration & All & All \\
\bottomrule
\end{tabularx}
\end{table}

\textbf{How the Tiers Work:}

\begin{enumerate}
\item \textbf{T0 (Baseline):} Start with an empty prompt (00-empty) to see what the raw model can do, then go all the way up to the full 1787-line CLAUDE.md (03-full). Individual blocks (B01-B18) let me test each piece of the prompt separately to see what actually matters.
\item \textbf{T1-T2 (Skills vs Tools):} T1 uses skills, domain knowledge baked into prompts, which is token-efficient. T2 uses external tools via JSON schemas, but loading all those tool definitions inflates token usage. I call this the "Token Efficiency Chasm", the gap between lean skill-based approaches and schema-heavy tool architectures.
\item \textbf{T3-T4 (Multi-Agent Setups):} T3 does flat delegation, breaking tasks into smaller pieces and assigning them to specialist agents. T4 adds hierarchy with self-correction loops, but this complexity can increase costs.
\item \textbf{T5 (Smart Combinations):} Take what works from the other tiers, combine them together in different combinations. A single test would have the best T1 skills, T2 tools, T3 agents, and T4 task delegation. We do not want to brute force here due to combinatorial explosion, but picking combinations of the top few categories can help give an idea of what combinations work best together.
\item \textbf{T6 (Everything):} Turn on everything at once. All skills, tools, agents, prompt segments, and servers. This I hope establishes the theoretical max performance and shows where diminishing returns kick in, but also can show signs of over-engineering if it is occurring.
\end{enumerate}

For each tier T(n), I compare it directly against T(n-1) to see what that
specific change actually achieves in terms of performance and cost. These
tiers map onto a broader multi-dimensional search space.

\subsection{Dimensional Search Space}\label{sec:dimensions}

The framework tests across four different dimensions. Each one is an independent
knob you can turn, and they all affect both what the agent can do and how much
it costs.

\subsubsection{Agent Complexity Axis (Tiers 0-6)}

The agent complexity axis spans from simple single-agent configurations to hierarchical multi-agent systems, as shown in Table~\ref{tab:agent-complexity}.

\begin{table}[htbp]
\centering
\caption{Agent Complexity Axis}
\label{tab:agent-complexity}
\begin{tabularx}{\textwidth}{llX}
\toprule
Tier Range & Complexity Level & Description \\
\midrule
T0 & Single-agent, prompt-only & Base model with varying prompt sophistication \\
T1 & Single-agent with skills & Add in agentic skills to improve the quality of the work \\
T2 & Single-agent with tools & External API access via tool schemas \\
T3 & Multi-agent, flat & Specialist agents with central orchestrator \\
T4 & Multi-agent, hierarchical & Nested orchestration with self-correction loops \\
T5 & Best case scenarios & Attempt to pick the best case scenarios from previous runs to see if the sum is more than its parts \\
T6 & Maximum configuration & All features enabled simultaneously \\
\bottomrule
\end{tabularx}
\end{table}

\subsubsection{Prompt Complexity Axis}

Prompt complexity is measured in lines of system prompt content, ranging from 0
(empty) to 1787 (full CLAUDE.md from ProjectOdyssey\cite{projectodyssey}), as detailed in Table~\ref{tab:prompt-complexity}.

\begin{table}[htbp]
\centering
\caption{Prompt Complexity Axis}
\label{tab:prompt-complexity}
\begin{tabular}{llll}
\toprule
Level & Lines & Description & Representative Test \\
\midrule
Empty & 0 & No system prompt & T0-00-empty \\
System & 0 & Only system prompt & T0-01-empty \\
Minimal & ~55 & Safety rules only & T0-06-B02 \\
Core & ~260 & Essential blocks (B02, B07, B18) & T0-03-core \\
Standard & ~400 & Seven core blocks & T0-02-standard \\
Full & 1787 & All 18 CLAUDE.md blocks & T0-03-full \\
\bottomrule
\end{tabular}
\end{table}

Each block (B01-B18) can be tested separately to see whether the part actually
contributes to the whole.

\subsubsection{Skill Complexity Axis}

Skills are organized by domain, as shown in Table~\ref{tab:skill-complexity}.

\begin{table}[htbp]
\centering
\caption{Skill Complexity Axis}
\label{tab:skill-complexity}
\begin{tabularx}{\textwidth}{llXl}
\toprule
Category & Count & Example Domains & Token Efficiency \\
\midrule
Agent & 5 & Agent management patterns & High \\
CI/CD & 7 & Build and deployment automation & High \\
Documentation & 4 & Technical writing assistance & Medium \\
GitHub & 10 & Repository management & Medium \\
Language & 10 & Programming language specific & High \\
Quality & 5 & Code quality and review & Medium \\
Workflow & 5 & Development workflow patterns & High \\
\bottomrule
\end{tabularx}
\end{table}

Table~\ref{tab:skill-complexity} shows the 46 skills across 7 categories tested in T1. Skills bake knowledge into prompts, so
you avoid loading massive tool schemas. But do these actually improve
performance? That is an open question. T6 enables all 61 available skills, including those tested individually in other tiers.

\subsubsection{Agent Hierarchy Axis}

Three ways to organize agents, tested across T3-T4:

\begin{table}[htbp]
\centering
\caption{Agent Hierarchy Axis}
\begin{tabularx}{\textwidth}{lXXX}
\toprule
Pattern & Coordination & Communication Overhead & Use Cases \\
\midrule
\textbf{Flat} & No supervision; peer-to-peer & Low & Simple, independent tasks \\
\textbf{Hierarchical} & L0-L4 levels with explicit supervision & High & Complex, interdependent tasks requiring planning \\
\textbf{Hybrid} & Selective hierarchy based on task complexity & Medium & Adaptive: flat for simple tasks, hierarchical for complex \\
\bottomrule
\end{tabularx}
\end{table}

Hierarchy matters for costs because each supervision layer adds more
orchestration tokens and potentially more self-correction iterations.

\section{Framework Architecture}\label{sec:architecture}

Having established what is tested (tiers T0-T6) and the multi-dimensional search space they explore, I now describe how the Scylla framework is built. The architecture is designed around four principles: model-agnostic evaluation via the adapter pattern, full reproducibility through git worktrees and checkpoint recovery, crash recovery with atomic state persistence, and parallelism at both tier and subtest levels to minimize evaluation time.

\subsection{System Overview}\label{sec:system-overview}

\begin{figure}[htbp]
\centering
\begin{tikzpicture}[
  layer/.style={draw, rectangle, rounded corners, minimum width=10cm,
    minimum height=1.2cm, text centered, font=\small, align=center},
  arrow/.style={-Stealth, thick}
]

\node[layer, fill=blue!10] (runner) at (0,0)
  {\textbf{E2E Runner} \\ {\scriptsize Experiment Orchestration}};

\node[layer, fill=blue!10, below=0.4cm of runner] (workspace)
  {\textbf{Workspace Manager} \\ {\scriptsize Git Worktrees + Isolation}};

\node[layer, fill=blue!10, below=0.4cm of workspace] (tier)
  {\textbf{Tier Manager} \\ {\scriptsize Config Injection (YAML/CLAUDE.md)}};

\node[layer, fill=blue!10, below=0.4cm of tier] (checkpoint)
  {\textbf{Checkpoint System} \\ {\scriptsize Resume + Rate Limiting}};

\node[layer, fill=green!10, below=0.4cm of checkpoint] (adapter)
  {\textbf{Adapter Layer} \\ {\scriptsize CLI Execution (Claude Code)}};

\node[layer, fill=orange!10, below=0.4cm of adapter] (judge)
  {\textbf{Judge Pipeline} \\ {\scriptsize LLM Scoring (Opus 4.5, Sonnet 4.5, Haiku 4.5)}};

\node[layer, fill=gray!10, below=0.4cm of judge] (analysis)
  {\textbf{Analysis Pipeline} \\ {\scriptsize Metrics Calculation + Visualization}};

\draw[arrow] (runner.south) -- (workspace.north);
\draw[arrow] (workspace.south) -- (tier.north);
\draw[arrow] (tier.south) -- (checkpoint.north);
\draw[arrow] (checkpoint.south) -- (adapter.north);
\draw[arrow] (adapter.south) -- (judge.north);
\draw[arrow] (judge.south) -- (analysis.north);

\end{tikzpicture}
\caption{System architecture showing the seven-layer pipeline. Data flows top-to-bottom through: orchestration, workspace isolation, configuration injection, checkpoint management, CLI execution, LLM evaluation, and metrics analysis.}
\label{fig:architecture}
\end{figure}
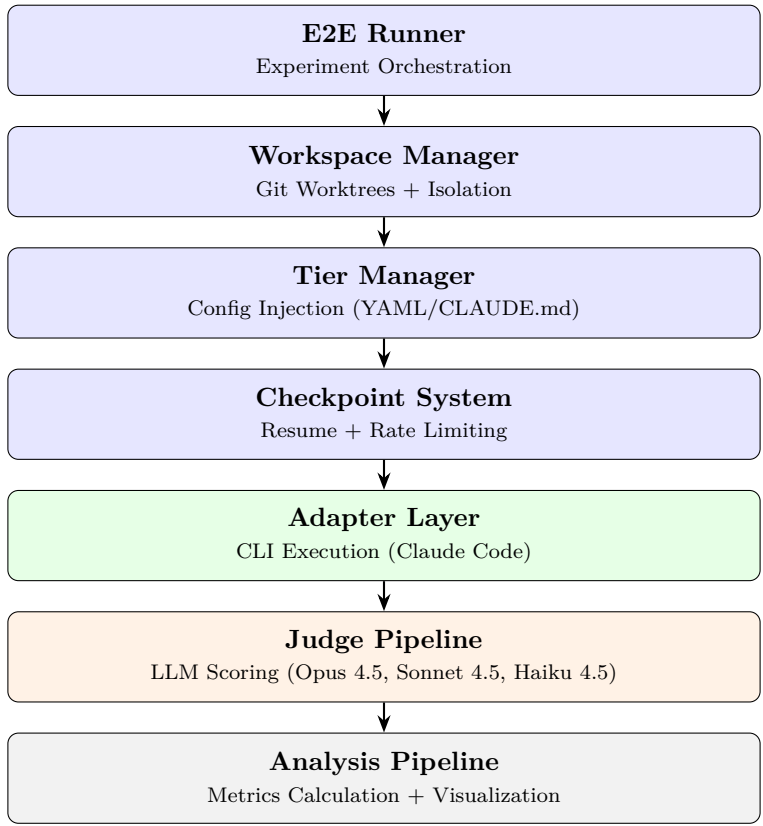

The framework consists of six major subsystems that coordinate to execute evaluations, collect artifacts, score results, and produce analyses. Figure~\ref{fig:architecture} shows the component hierarchy and data flow.

The E2E Runner coordinates all evaluation steps, managing parallel execution of tiers and subtests while enforcing dependency constraints (T5 waits for T0-T4; T6 waits for T5). The Workspace Manager creates isolated git worktrees at pinned commits to prevent cross-contamination. The Tier Manager injects configuration files (CLAUDE.md, skills, agent definitions) via symlinks into each workspace. The Checkpoint System saves JSON checkpoints after every completed run, enabling crash recovery and rate limit coordination. The Adapter Layer executes CLI commands via a standardized interface (AdapterConfig → AdapterResult). The Judge Pipeline runs three LLM judges sequentially, scoring outputs using weighted rubrics. The Analysis Pipeline aggregates data into DataFrames and generates figures and tables.

\subsection{Adapter Layer}\label{sec:adapter}

The adapter pattern is how Scylla achieves model-agnostic evaluation. All agent implementations inherit from the \texttt{BaseAdapter} abstract class, which defines two methods: \texttt{execute(config: AdapterConfig) -> AdapterResult} for running evaluations, and \texttt{extract\_tokens(stdout: str, stderr: str) -> AdapterTokenStats} for parsing token usage from CLI output.

The current implementation includes the \texttt{ClaudeCodeAdapter} for Claude Code CLI, which is used for all evaluations in this paper. Each adapter receives an \texttt{AdapterConfig} containing the model, prompt file, workspace path, output directory, timeout, and environment variables. The adapter executes the CLI command, captures stdout/stderr, extracts token statistics from the output, calculates costs using vendor pricing tables, and returns an \texttt{AdapterResult} with exit code, output, duration, token stats, cost, API call count, and timeout status.

This design enables future cross-vendor comparisons using identical tier configurations. Additional adapters for tools like Cline, OpenAI Codex, or other CLI-based coding assistants can be implemented by inheriting from \texttt{BaseAdapter} and providing tool-specific execution and token extraction logic. Token extraction logic adapts to each vendor's output format, but the framework sees a standardized \texttt{AdapterTokenStats} regardless of source.

\subsection{Execution Pipeline}\label{sec:exec-pipeline}

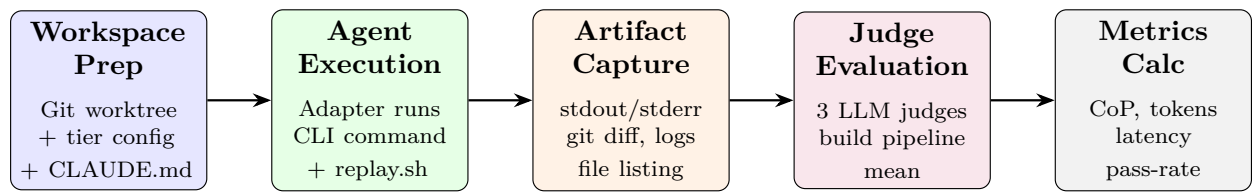
\begin{figure}[htbp]
\centering
\resizebox{\textwidth}{!}{%
\begin{tikzpicture}[
  block/.style={draw, rectangle, rounded corners, text width=2.2cm,
    minimum height=1.5cm, text centered, font=\small, align=center},
  arrow/.style={-Stealth, thick}
]

\node[block, fill=blue!10] (workspace) at (0,0)
  {\textbf{Workspace Prep} \\ \vspace{2mm} {\scriptsize Git worktree \\ + tier config \\ + CLAUDE.md}};
\node[block, fill=green!10, right=0.8cm of workspace] (agent)
  {\textbf{Agent Execution} \\ \vspace{2mm} {\scriptsize Adapter runs \\ CLI command \\ + replay.sh}};
\node[block, fill=orange!10, right=0.8cm of agent] (artifact)
  {\textbf{Artifact Capture} \\ \vspace{2mm} {\scriptsize stdout/stderr \\ git diff, logs \\ file listing}};
\node[block, fill=purple!10, right=0.8cm of artifact] (judge)
  {\textbf{Judge Evaluation} \\ \vspace{2mm} {\scriptsize 3 LLM judges \\ build pipeline \\ mean}};
\node[block, fill=gray!10, right=0.8cm of judge] (metrics)
  {\textbf{Metrics Calc} \\ \vspace{2mm} {\scriptsize CoP, tokens \\ latency \\ pass-rate}};

\draw[arrow] (workspace) -- (agent);
\draw[arrow] (agent) -- (artifact);
\draw[arrow] (artifact) -- (judge);
\draw[arrow] (judge) -- (metrics);

\end{tikzpicture}%
}
\caption{Execution pipeline for a single evaluation run. Each step produces inputs for the next, enabling reproducibility through saved artifacts at every boundary.}
\label{fig:exec-pipeline}
\end{figure}

Every evaluation run follows a five-step pipeline from workspace preparation to metrics calculation. Figure~\ref{fig:exec-pipeline} shows the sequential flow.

\textbf{Step 1: Workspace Prep} creates a fresh git worktree at the pinned commit and injects tier-specific configuration files via symlinks:

\begin{itemize}
\item CLAUDE.md from \texttt{config/tiers/TN/subtest-NN/CLAUDE.md}
\item Skills from \texttt{config/tiers/TN/subtest-NN/.claude-plugin/skills}
\item Agents from \texttt{config/tiers/TN/subtest-NN/.claude/agents}
\item Generates a \texttt{replay.sh} script for manual reproduction
\end{itemize}

\textbf{Step 2: Agent Execution} invokes the adapter's \texttt{execute()} method with the workspace path and prompt file. The adapter runs the CLI command (e.g., \texttt{claude-code < prompt.txt}) with a timeout (default 3600s), captures stdout/stderr via subprocess pipes, and saves the output to \texttt{output\_dir/stdout.txt} and \texttt{output\_dir/stderr.txt}.

\textbf{Step 3: Artifact Capture} collects all execution outputs: stdout/stderr, git diff of file changes, complete file listing, CLI logs, and the generated \texttt{replay.sh} script. Artifacts are saved to \texttt{results/test-NNN/TN/subtest-NN/run-MMMMM/} for later analysis.

\textbf{Step 4: Judge Evaluation} runs three LLM judges sequentially (Opus 4.5, Sonnet 4.5, Haiku 4.5). Each judge receives the same prompt containing the task description, expected output, rubric, and captured artifacts. Judges score five weighted categories (functional correctness, code quality, proportionality, build pipeline, overall quality) and return a final score and grade. The framework computes the mean of the three scores for consensus.

\textbf{Step 5: Metrics Calculation} aggregates judge scores, token statistics, and timing data to compute Pass-Rate, Implementation Rate, Cost-of-Pass, token distribution, latency breakdown, and judge agreement metrics. Results are saved to \texttt{runs.csv}, \texttt{judges.csv}, \texttt{criteria.csv}, and \texttt{summary.json}.

\subsection{Tier Dependencies and Parallelism}\label{sec:tier-deps}

\begin{figure}[htbp]
\centering
\begin{tikzpicture}[
  tier/.style={draw, rectangle, rounded corners, minimum width=1.2cm,
    minimum height=0.8cm, text centered, font=\small},
  arrow/.style={-Stealth, thick},
  phase/.style={draw, dashed, rounded corners, inner sep=8pt, font=\scriptsize}
]

\node[tier, fill=blue!10] (t0) at (0,0) {T0};
\node[tier, fill=blue!10, below=0.3cm of t0] (t1) {T1};
\node[tier, fill=blue!10, below=0.3cm of t1] (t2) {T2};
\node[tier, fill=blue!10, below=0.3cm of t2] (t3) {T3};
\node[tier, fill=blue!10, below=0.3cm of t3] (t4) {T4};

\node[tier, fill=green!10] (t5) at (4,-2) {T5};

\node[tier, fill=orange!10] (t6) at (7,-2) {T6};

\draw[arrow] (t0.east) -- (t5.west);
\draw[arrow] (t1.east) -- (t5.west);
\draw[arrow] (t2.east) -- (t5.west);
\draw[arrow] (t3.east) -- (t5.west);
\draw[arrow] (t4.east) -- (t5.west);

\draw[arrow] (t5) -- (t6);

\node[phase, fit=(t0)(t4), label=above:\textbf{Phase 1}] (p1) {};
\node[phase, fit=(t5), label=above:\textbf{Phase 2}] (p2) {};
\node[phase, fit=(t6), label=above:\textbf{Phase 3}] (p3) {};

\node[font=\tiny, below=0.2cm of p1] {parallel};
\node[font=\tiny, below=0.2cm of p2] {merge best};
\node[font=\tiny, below=0.2cm of p3] {all};

\end{tikzpicture}
\caption{Tier dependency graph and parallel execution model. T0-T4 execute in parallel (Phase 1). T5 waits for T0-T4 completion and merges their best configurations (Phase 2). T6 waits for T5 and enables all features (Phase 3).}
\label{fig:tier-deps}
\end{figure}
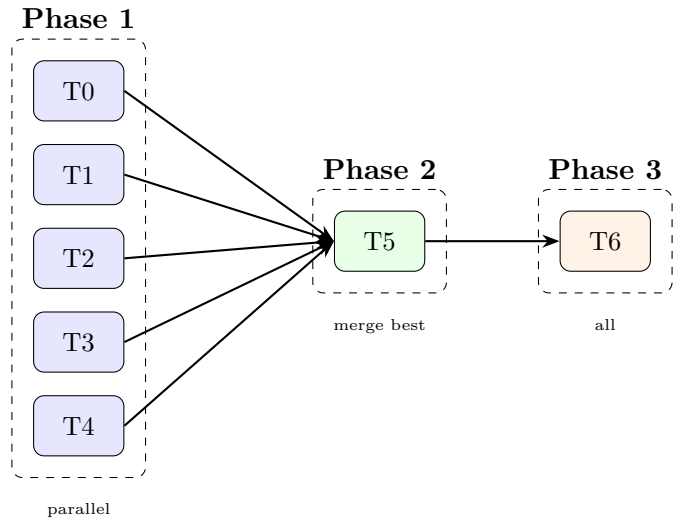

Tiers execute in three sequential phases based on dependency constraints. Figure~\ref{fig:tier-deps} visualizes the dependency graph.

\textbf{Phase 1 (Parallel)}: T0-T4 have no dependencies and execute in parallel using a \texttt{ThreadPoolExecutor}. Each tier runs all its subtests concurrently via a \texttt{ProcessPoolExecutor} for subtest-level parallelism. This reduces total evaluation time from 7 sequential tier runs to 3 sequential phases with internal parallelism.

\textbf{Phase 2 (Merge)}: T5 depends on T0-T4 and cannot start until all Phase 1 tiers complete. The Tier Manager merges configurations using the following rules: skills are unioned across all tiers (if T1 uses skill A and T3 uses skill B, T5 gets both), agents are unioned similarly, blocks from the highest-scoring tier replace conflicts (if T0 and T2 both define block B02, T5 uses the version from whichever tier scored higher), and tools follow an "all wins" policy (if any tier enables tools, T5 enables them).

\textbf{Phase 3 (Everything)}: T6 depends on T5 and enables maximum configuration: all 61 skills, all tools, all 44 agents, and all CLAUDE.md blocks. This establishes the theoretical performance ceiling and demonstrates diminishing returns from over-engineering (as shown in Results).

Parallelism is bounded by a global semaphore limiting concurrent agent executions to prevent API rate limit violations. The Checkpoint System coordinates rate limits across parallel workers by tracking timestamps and enforcing minimum delays between API calls.

\subsection{Checkpoint and Reproducibility}\label{sec:checkpoint}

Scylla is designed for long-running experiments that may crash or hit rate limits. The Checkpoint System saves progress after every completed run, enabling seamless resume.

After each run completes, the runner writes a checkpoint JSON to {\small\texttt{results/test-NNN/checkpoint.json.tmp}}, atomically renames it to \texttt{checkpoint.json}, and validates the configuration hash on resume to detect experiment tampering. The checkpoint contains the list of completed runs (tier, subtest, run\_id, timestamp, pass/fail), current rate limit state (last API call timestamp, tokens consumed), and experiment metadata (test ID, total runs planned, start time).

On resume, the runner reads the checkpoint, validates the config hash matches the current experiment, skips already-completed runs, and continues from the next pending run. This allows experiments to survive crashes, API timeouts, network failures, and manual interruptions (Ctrl+C) without losing work.

Reproducibility is further ensured by the \texttt{replay.sh} script generated for each run. The script contains the exact git commit, CLI command, environment variables, and workspace setup needed to manually reproduce the run. This enables debugging, manual inspection, and verification of framework behavior.

Having described how the framework is built, I now turn to the specific metrics used to quantify performance, quality, and economic trade-offs.

\section{Test Metrics}\label{sec:metrics}

\subsection{Performance Metrics}

\textbf{Pass-Rate} is straightforward, did it work or not:

$$\text{Pass-Rate} = \frac{\text{correct\_solutions}}{\text{total\_attempts}}$$

Range is 0.0 (nothing worked) to 1.0 (everything worked). "Correct" means it
passes the test suite for that specific task. Report this with confidence
 intervals (95\% CI when sample sizes exceed 30).

\textbf{Fine-Grained Progress Rate} ($R_{Prog}$) tracks how far you got through
 multi-step tasks:

$$R_{Prog} = \frac{\text{achieved\_progress\_steps}}{\text{expected\_progress\_steps}}$$

Range is 0.0 to 1.0. If you get 1.0, it means the agent completed all expected steps. This is particularly useful for debugging where things go wrong
in complex workflows, especially in hierarchical setups with all their
self-correction loops.

\textbf{Consistency} measures how stable the outputs are:

$$\text{Consistency} = 1 - \frac{\sigma(\text{outputs})}{\mu(\text{outputs})}$$

Range is 0.0 to 1.0, higher means more deterministic. Matters most for where
you are trying to get reliable structured outputs.

\subsection{Quality Metrics}

\textbf{Implementation Rate} (Impl-Rate) measures whether you actually satisfied the
requirements:

$$\text{Impl-Rate} = \frac{\text{satisfied\_requirements}}{\text{total\_requirements}}$$

Range is 0.0 to 1.0. This gives you more detail than just pass/fail, you get
partial credit for incomplete work. Checked using multiple LLM judges with
mean scoring for consensus.

\subsection{Efficiency and Cost Metrics}

\textbf{Latency} is just time from start to finish (seconds):

\begin{itemize}
\item Time-to-First-Token (TTFT)
\item Total response time
\item Tool execution time

\end{itemize}

It matters a lot for architectures where verification loops can really slow
things down.

\textbf{Token Distribution} shows where your tokens are going:

$$\text{token\_dist} = \left\{ \frac{\text{input\_tokens}}{\text{total\_tokens}}, \frac{\text{output\_tokens}}{\text{total\_tokens}}, \frac{\text{tool\_input\_tokens}}{\text{total\_tokens}}, \frac{\text{tool\_output\_tokens}}{\text{total\_tokens}} \right\}$$

Useful for identifying what is actually contributing to the cost (like T3's
massive agent prompts or T4's orchestration overhead).

\textbf{Cost-of-Pass (CoP)} is the primary metric, what is the expected cost to get
one correct solution:

$$\text{CoP} = \frac{\text{total\_cost}}{\text{pass\_rate}}$$

Units are USD. Lower is better. If pass-rate hits zero, CoP goes to infinity,
that configuration is economically dead. This combines both cost and accuracy
into one number that tells you if something is actually sustainable.

\textbf{Frontier CoP} represents the best CoP for all the various tests:

$$\text{Frontier\_CoP} = \min(\text{CoP}_{T0}, \text{CoP}_{T1}, \ldots, \text{CoP}_{T6})$$

This metric currently is just the minimum CoP across all tiers. Comparing this
against what it costs to hire a human expert will allow developers to see if
automation actually makes economic sense. Different model providers will have
different cost assumptions.

\begin{table}[htbp]
\centering
\caption{Model Pricing (as of January 2026)}
\begin{tabular}{lll}
\toprule
Model & Input (\$/1M tokens) & Output (\$/1M tokens) \\
\midrule
Claude Opus 4.5 & \$15.00 & \$75.00 \\
Claude Sonnet 4.5 & \$3.00 & \$15.00 \\
Claude Haiku 4.5 & \$1.00 & \$5.00 \\
\bottomrule
\end{tabular}
\end{table}

\section{Test Configuration}

\subsection{Hardware and Infrastructure}

\begin{table}[htbp]
\centering
\caption{Hardware and Infrastructure}
\begin{tabular}{ll}
\toprule
Component & Specification \\
\midrule
Platform & Linux (WSL2) \\
Kernel & 6.6.87.2-microsoft-standard-WSL2 \\
Isolation & Each test runs in clean workspace \\
Compute & Standard CPU (no GPU required for evaluation) \\
\bottomrule
\end{tabular}
\end{table}

Each test runs in its own git worktree with the repo at a specific git
commit. This means every run is reproducible and tests cannot mess with each
other. Every worktree starts fresh with:

\begin{itemize}
\item Clean git workspace at the exact commit specified
\item Tier-specific config files
\item Whatever tools/skills that tier needs
\item Isolated filesystem for collecting results
\end{itemize}

\subsection{Software Stack}

\begin{table}[htbp]
\centering
\caption{Software Stack}
\begin{tabular}{ll}
\toprule
Component & Version/Tool \\
\midrule
CLI Tool & Claude Code\footnote{Claude Code requires either an Anthropic API key or a Claude Max subscription for operation.} (primary evaluation target) \\
Language Runtime & Python 3.12.3, Mojo 0.26.1.0.dev2025122805 (211e2f5c) \\
Package Manager & Pixi \\
Isolation & Git Worktrees \\
Orchestration & Custom Scylla framework (Section~\ref{sec:architecture}) \\
Validation & JSON Schema, YAML validation \\
Version Control & Git Version 2.43.0 \\
\bottomrule
\end{tabular}
\end{table}

The evaluation harness does five things:

\begin{enumerate}
\item \textbf{Workspace Prep}: Clone the repo, check out the specific commit, inject tier config
\item \textbf{Run the Agent}: Fire up Claude Code with whatever prompt/tools that tier uses
\item \textbf{Capture Everything}: Grab the output, command logs, file changes, artifacts
\item \textbf{Judge It}: Run three LLM judges sequentially (Opus, Sonnet, Haiku)
\item \textbf{Calculate Metrics}: Crunch the numbers for Pass-Rate, Impl-Rate, CoP, token usage, consensus scores
\end{enumerate}

\subsection{Model Configuration}

\begin{table}[htbp]
\centering
\caption{Execution Models (performing the tasks)}
\begin{tabular}{lll}
\toprule
Model & Model ID & Primary Use \\
\midrule
Claude Opus 4.5 & claude-opus-4-5-20251101 & complex reasoning, hierarchical orchestration \\
Claude Sonnet 4.5 & claude-sonnet-4-5-20250929 & standard execution, balanced cost/capability \\
Claude Haiku 4.5 & claude-haiku-4-5-20251001 & simple tasks, cost optimization \\
\bottomrule
\end{tabular}
\end{table}

\textbf{Judge Configuration} (evaluating the outputs):

\begin{itemize}
\item Three judges per evaluation: Opus 4.5, Sonnet 4.5, Haiku 4.5
\item Take the mean of the three scores for consensus
\item Same prompt for all judges (only the model changes)
\item Judge prompt: \texttt{config/judge/system\_prompt.md}
\end{itemize}

\textbf{Safety}:

\begin{itemize}
\item Safety-net plugin blocks destructive operations
\end{itemize}

\textbf{Model-Agnostic Framework Design}:

The framework is designed to work with any CLI tool or model through standardized
interfaces. Everything goes through the CLI's language interface and filesystem
outputs without touching model APIs directly. This enables consistent metrics
(CoP, Pass-Rate, Impl-Rate) across all models for apples-to-apples economic
comparisons. The tier structure (T0-T6) applies to all tools, allowing direct
architectural comparisons across vendors. Everything is in version-controlled
YAML (model IDs, temperature, token limits), making it easy to reproduce across
different tools and swap judges. The adapter pattern enabling this is described in Section~\ref{sec:adapter}.

\section{Test Cases}

\subsection{Pull Request (PR) Selection Criteria}

Test cases come from real software development tasks. Here is what I consider to
make a good test:

\begin{enumerate}
\item \textbf{Reproducible}: Pin it to a specific git commit
\item \textbf{Clear success criteria}: Can be expressed in a rubric with measurable requirements
\item \textbf{Representative}: Real work that developers actually do
\item \textbf{Incrementally complex}: From trivial (Hello World) to multi-file architecture changes
\item \textbf{Unambiguous}: Clear task, clear expected outcome
\end{enumerate}

\begin{table}[htbp]
\centering
\caption{Size Categories}
\begin{tabularx}{\textwidth}{llXX}
\toprule
Category & Lines of Code (LOC) & Complexity Characteristics & Example Tasks \\
\midrule
\textbf{Small} & < 100 LOC & Single file changes, configuration updates & Config file modification, simple script creation \\
\textbf{Medium} & 100-500 LOC & Feature additions, localized refactoring & Add validation logic, implement utility function \\
\textbf{Large} & 500-2000 LOC & Multi-file features, architectural changes & New module implementation, build system migration \\
\bottomrule
\end{tabularx}
\end{table}

Complexity also depends on:
\begin{itemize}
\item How many tool calls you need
\item How much of the codebase you have to understand
\item How many sequential steps
\item How many constraints you are working under
\end{itemize}

\subsection{Workflow Categories}

Different categories test different capabilities:

\begin{table}[htbp]
\centering
\caption{Workflow Categories}
\begin{tabularx}{\textwidth}{lXlX}
\toprule
Category & Description & Complexity & Key Challenges \\
\midrule
\textbf{Build System} & Makefile, Justfile, build automation configuration & Low-Medium & Syntax correctness, equivalence preservation \\
\textbf{CI/CD} & GitHub Actions, deployment pipelines, automation & Medium & Multi-file coordination, environment configuration \\
\textbf{Bug Fixing} & Defect resolution from issue description & Medium-High & Root cause diagnosis, minimal change principle \\
\textbf{New Features} & Feature implementation from requirements & High & Requirements interpretation, design decisions \\
\textbf{Refactoring} & Code restructuring without behavior change & Medium & Behavior preservation, test coverage \\
\textbf{Optimization} & Performance improvements, algorithmic enhancements & Medium-High & Profiling, benchmarking, trade-off analysis \\
\textbf{Review} & Code review and feedback generation & Medium & Pattern recognition, best practice knowledge \\
\textbf{Documentation} & Technical documentation generation & Low-Medium & Clarity, completeness, accuracy \\
\textbf{Issue Filing} & Bug report creation from symptoms & Low & Information gathering, reproduction steps \\
\bottomrule
\end{tabularx}
\end{table}

\subsection{Test Case Matrix}

I have designed \textbf{47 planned test cases} spanning five complexity bands (baseline validation, build system tasks, feature implementation, bug fixing/refactoring, complex multi-step tasks). Each test is defined in YAML with pinned repository commits, task prompts, validation rubrics, and tier configurations. Tests are structured to increase in difficulty progressively. See Section~\ref{sec:further} (Further Work) for planned full-scale execution.

\section{Results}\label{sec:results}

I will present results from the dryrun experiment (test-001, Hello World task)
across all seven tiers. The dryrun serves as a pipeline validation exercise with
N=1 run per tier, establishing that the framework executes end-to-end
successfully and generates the expected metrics, figures, and tables. Think of
this as a "smoke test", if the pipeline works on the simplest possible task, I
know it will handle the complex stuff later.

\subsection{Pipeline Validation (dryrun Overview)}

First, the dryrun was executed with the following setup:

\begin{itemize}
\item \textbf{Scope}: 1 model (Sonnet 4.5), 7 tiers (T0-T6), 1 subtest per tier
\item \textbf{Judges}: 3 judges per run (Opus 4.5, Sonnet 4.5, Haiku 4.5) = 21 total judge evaluations
\item \textbf{Criteria}: 5 criteria per judge $\times$ 21 judges = 105 total criteria scores
\item \textbf{Total cost}: \$1.01 (agent execution + judge evaluation)
\item \textbf{Total duration}: ~1289 seconds (~21.5 minutes) sum of per-tier durations; actual wall-clock time was ~550 seconds due to parallel execution
\item \textbf{Pass rate}: 100\% (all 7 tiers passed, all grade A)
\end{itemize}

Table~\ref{tab:tier-summary} shows the tier-by-tier summary. All tiers achieved grade A with mean
consensus scores ranging from 0.943 (T6) to 0.983 (T2, T3, T5). The task is
trivially easy, as expected, even T0 (minimal prompt) scores 0.973, with T4 at 0.960.

\begin{table}[htbp]
\centering
\caption{Tier Summary (dryrun)}\label{tab:tier-summary}
\begin{tabular}{lllll}
\toprule
Tier & Pass Rate & Mean Score & Grade & CoP (\$) \\
\midrule
T0 & 1.000 & 0.973 & A & 0.135 \\
T1 & 1.000 & 0.970 & A & 0.127 \\
T2 & 1.000 & 0.983 & A & 0.138 \\
T3 & 1.000 & 0.983 & A & 0.129 \\
T4 & 1.000 & 0.960 & A & 0.168 \\
T5 & 1.000 & 0.983 & A & 0.065 \\
T6 & 1.000 & 0.943 & A & 0.247 \\
\bottomrule
\end{tabular}
\end{table}

\textbf{Key finding}: Quality converges across all tiers (ceiling effect), but cost
varies 3.8$\times$ from \$0.065 to \$0.247.

These results set the stage for deeper economic analysis of how architectural choices affect cost without improving quality on ceiling-constrained tasks.

\subsection{Cost-of-Pass Analysis}

Since all tiers pass (pass-rate = 1.0), Cost-of-Pass equals the raw cost. The Frontier CoP is \$0.065 (achieved by T5 hybrid).

\textbf{Cost ranking} (lowest to highest):
\begin{enumerate}
\item \textbf{T5} (hybrid): \$0.065 ,  Frontier CoP achieved through selective skill loading and minimal cache creation (4.6K vs 23-44K for other tiers)
\item \textbf{T1} (skills): \$0.127 ,  Token-efficient skill-based approach
\item \textbf{T3} (delegation): \$0.129 ,  Flat multi-agent with efficient orchestration
\item \textbf{T0} (baseline): \$0.135 ,  Minimal prompt overhead
\item \textbf{T2} (tooling): \$0.138 ,  Tool schema loading increases cache tokens
\item \textbf{T4} (hierarchy): \$0.168 ,  Hierarchical orchestration adds 30\% overhead vs T3
\item \textbf{T6} (super): \$0.247 ,  Maximum configuration; diminishing returns evident.
\end{enumerate}

T6 (everything enabled) costs the most despite scoring the lowest (0.943). This
is a maximalist approach, to see when more equals better.

\subsection{Token Analysis}\label{sec:token-analysis}

\begin{figure}[htbp]
\centering
\includegraphics[width=0.9\textwidth]{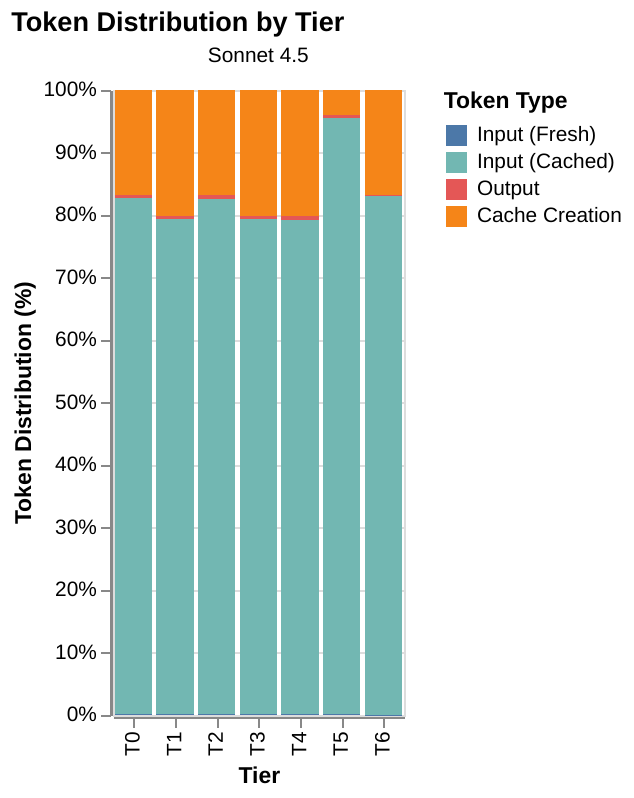}
\caption{Token distribution by tier and type. Stacked bar chart showing the breakdown of input, output, cache create, and cache read tokens across T0--T6. Cache read tokens dominate (79--95\%), consistent with prompt caching efficacy. However, T6's 218K cache reads versus T0's 113K illustrate the Token Efficiency Chasm, where architectural enhancements double token consumption without quality gains.}
\label{fig:fig07_token_distribution}
\end{figure}

Token distribution reveals where costs originate. Figure~\ref{fig:fig07_token_distribution} shows the breakdown by token type.

Cache read tokens dominate, ~79--95\% of total tokens across tiers (79--83\% excluding T5's 95\% outlier), showing
prompt caching works. But cache creation tokens vary dramatically:

\begin{table}[htbp]
\centering
\caption{Token Breakdown}
\begin{tabular}{lrrrrr}
\toprule
Tier & Input & Output & Cache Create & Cache Read & Total \\
\midrule
T0 & 29 & 656 & 23,106 & 112,686 & 136,477 \\
T1 & 25 & 558 & 23,266 & 91,477 & 115,326 \\
T2 & 29 & 711 & 23,350 & 113,858 & 137,948 \\
T3 & 25 & 668 & 23,352 & 91,771 & 115,816 \\
T4 & 23 & 725 & 23,556 & 91,828 & 116,132 \\
T5 & 26 & 625 & \textbf{4,629} & 109,368 & 114,648 \\
T6 & 29 & 722 & \textbf{44,337} & 218,778 & 263,866 \\
\bottomrule
\end{tabular}
\end{table}

The Token Efficiency Chasm mentioned in Section~\ref{sec:tiered-ablation} is supported by this data. T6 requires 218K cache read tokens versus T0's 113K, a 1.94x
increase (nearly double). T5 achieves efficiency by minimizing cache creation
(4.6K vs 23-44K), supporting the hybrid strategy.

Output tokens stay stable at 558-725 across tiers, showing the task itself
requires similar generation regardless of architecture.

\subsection{Latency Analysis}

Latency breaks into two components: agent execution time and judge evaluation
time, as shown in Table~\ref{tab:latency-breakdown}.

\begin{table}[htbp]
\centering
\caption{Latency Breakdown}
\label{tab:latency-breakdown}
\begin{tabular}{lrrrr}
\toprule
Tier & Agent Time (s) & Judge Time (s) & Total Time (s) & Judge \% of Total \\
\midrule
T0 & 35.3 & 167.8 & 203.1 & 82.6\% \\
T1 & 29.3 & 178.0 & 207.3 & 85.9\% \\
T2 & 36.8 & 161.7 & 198.5 & 81.5\% \\
T3 & 29.9 & 149.1 & 179.0 & 83.3\% \\
T4 & 41.2 & 137.0 & 178.2 & 76.9\% \\
T5 & 24.8 & 128.4 & 153.1 & 83.8\% \\
T6 & 28.4 & 141.1 & 169.5 & 83.2\% \\
\bottomrule
\end{tabular}
\end{table}

Judge evaluation dominates, 77-86\% of total latency, ranging from 128-178
seconds. This makes sense since 3 judges each evaluate the output independently.

Agent time varies modestly, 24.8-41.2 seconds. T5 is fastest (24.8s), T4 slowest
(41.2s). T5's speed advantage aligns with its cost advantage, both stem from
minimal cache loading.

On this trivial task, judge overhead dwarfs agent execution time, since there
are three judges for this simple task. On more complex tasks with multi-step
reasoning, agent time would dominate.

The latency patterns raise questions about judge consensus: how consistent are the three judges, and does the multi-judge design provide reliable scoring?

\subsection{Judge Agreement}

\begin{figure}[htbp]
\centering
\includegraphics[width=0.9\textwidth]{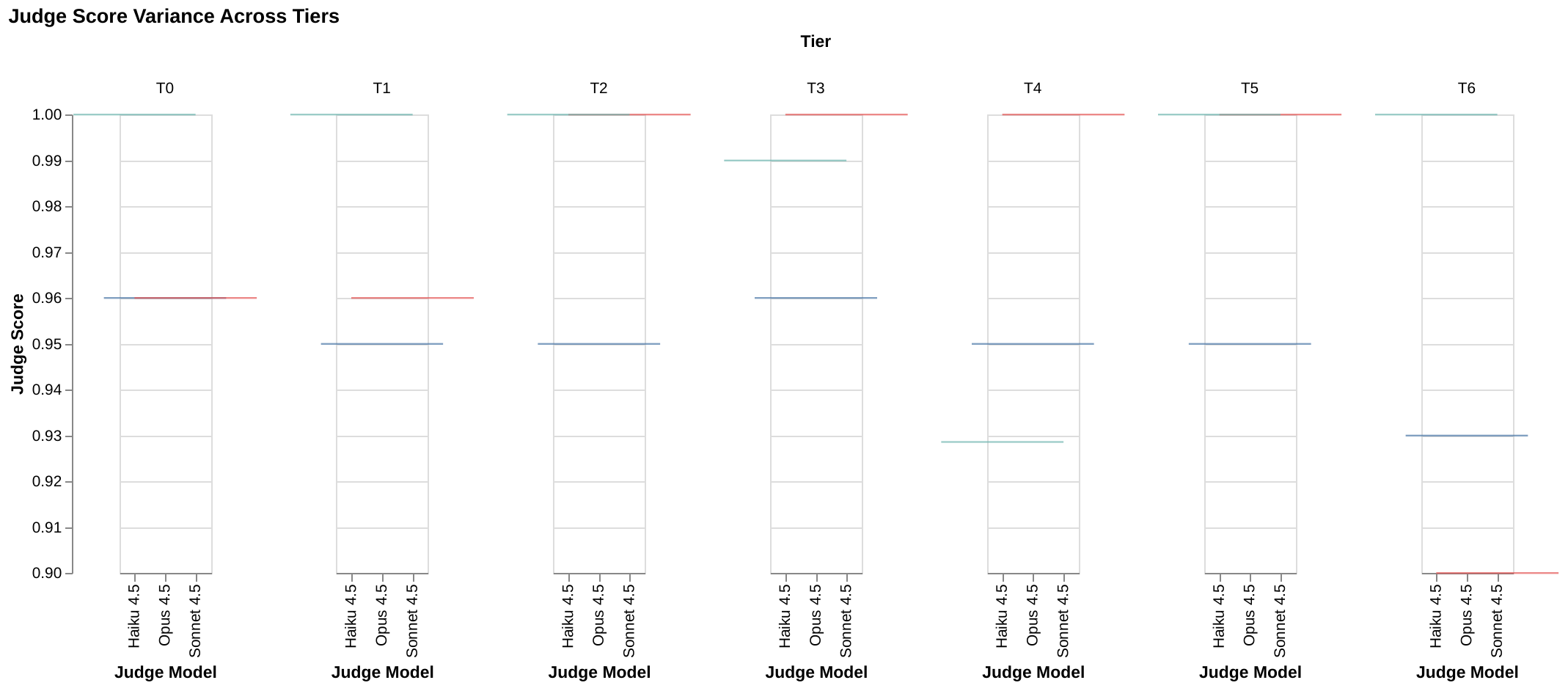}
\caption{Per-judge scoring variance across tiers. Box plots showing score distributions for each judge model (Opus 4.5, Sonnet 4.5, Haiku 4.5) faceted by tier. Opus exhibits the tightest distribution (most conservative), Haiku the widest (most generous), revealing systematic inter-judge bias that affects aggregate score reliability.}
\label{fig:fig02_judge_variance}
\end{figure}

\begin{figure}[htbp]
\centering
\includegraphics[width=0.9\textwidth]{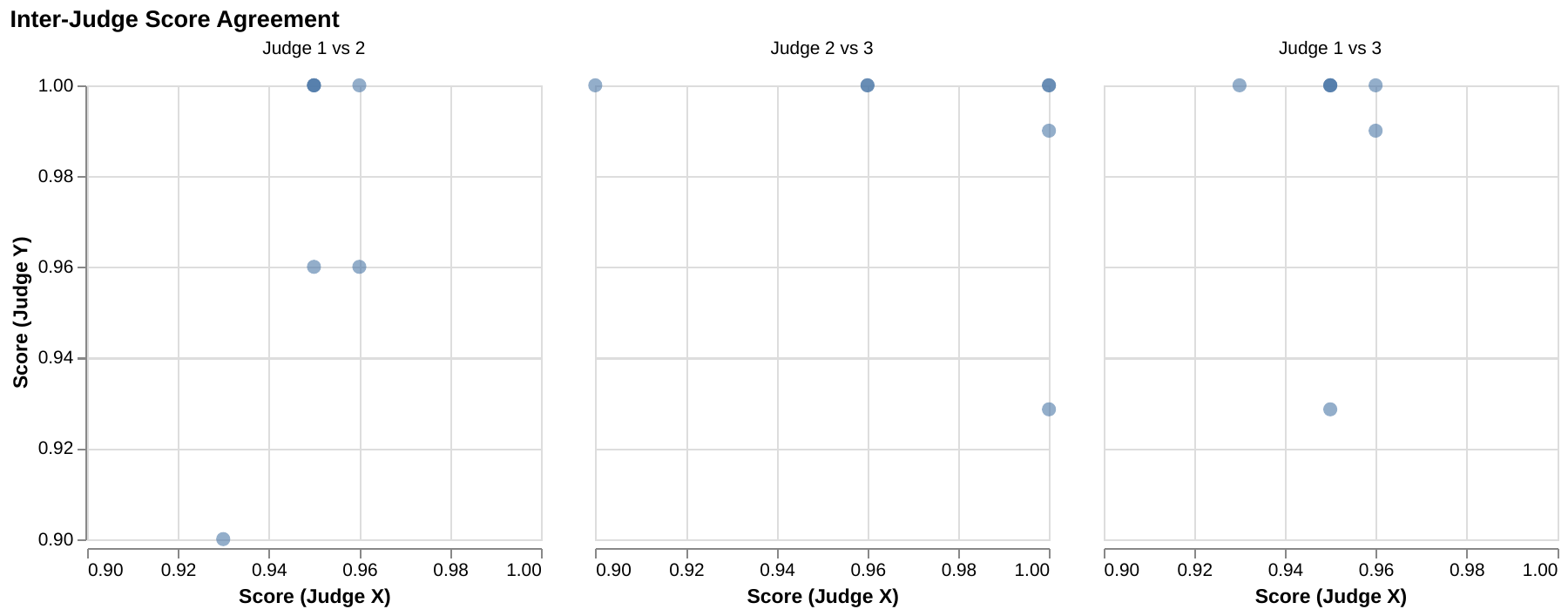}
\caption{Inter-judge agreement scatter matrix. Pairwise scatter plots showing score correlations between all judge pairs (Opus-Sonnet, Opus-Haiku, Sonnet-Haiku) with Spearman and Pearson correlation coefficients. Low-to-moderate correlations reveal systematic bias between judges rather than strong agreement, with Opus-Sonnet showing the highest concordance (Pearson r=0.706).}
\label{fig:fig14_judge_agreement}
\end{figure}

Three judges (Opus 4.5, Sonnet 4.5, Haiku 4.5) evaluated each run. Figure~\ref{fig:fig02_judge_variance} and Figure~\ref{fig:fig14_judge_agreement} show judge variance and pairwise agreement.

\textbf{Judge behavior patterns}:
\begin{itemize}
\item \textbf{Opus}: Most conservative judge, scores range 0.93-0.96, never awards S grade
\item \textbf{Sonnet}: Moderate judge, scores range 0.90-1.00, awards S grade in 4/7 tiers (T2, T3, T4, T5)
\item \textbf{Haiku}: Most generous judge, scores range 0.929-1.00, awards S grade in 5/7 tiers
\end{itemize}

\textbf{Pairwise agreement}:
\begin{itemize}
\item \textbf{Opus-Sonnet}: Spearman $\rho$ = 0.333, Pearson r = 0.706, mean $\Delta$ = 0.033
\item \textbf{Opus-Haiku}: Spearman $\rho$ = -0.273, Pearson r = -0.063, mean $\Delta$ = 0.045
\item \textbf{Sonnet-Haiku}: Spearman $\rho$ = -0.522, Pearson r = -0.347, mean $\Delta$ = 0.037
\end{itemize}

Krippendorff's $\alpha$ (interval): -0.117. Poor agreement, but expected with N=1 per
tier. \textbf{Note}: N=7 is insufficient for reliable correlation estimates; these
values are reported for completeness but interpret them with extreme caution.

Despite low inter-rater agreement, the 3-judge mean produces stable final
scores. The mean balances extreme scores, Haiku's 1.00 perfects versus Opus's
0.93 conservatism. This supports the multi-judge consensus design.

\subsection{Criteria Breakdown}

\begin{figure}[htbp]
\centering
\includegraphics[width=0.9\textwidth]{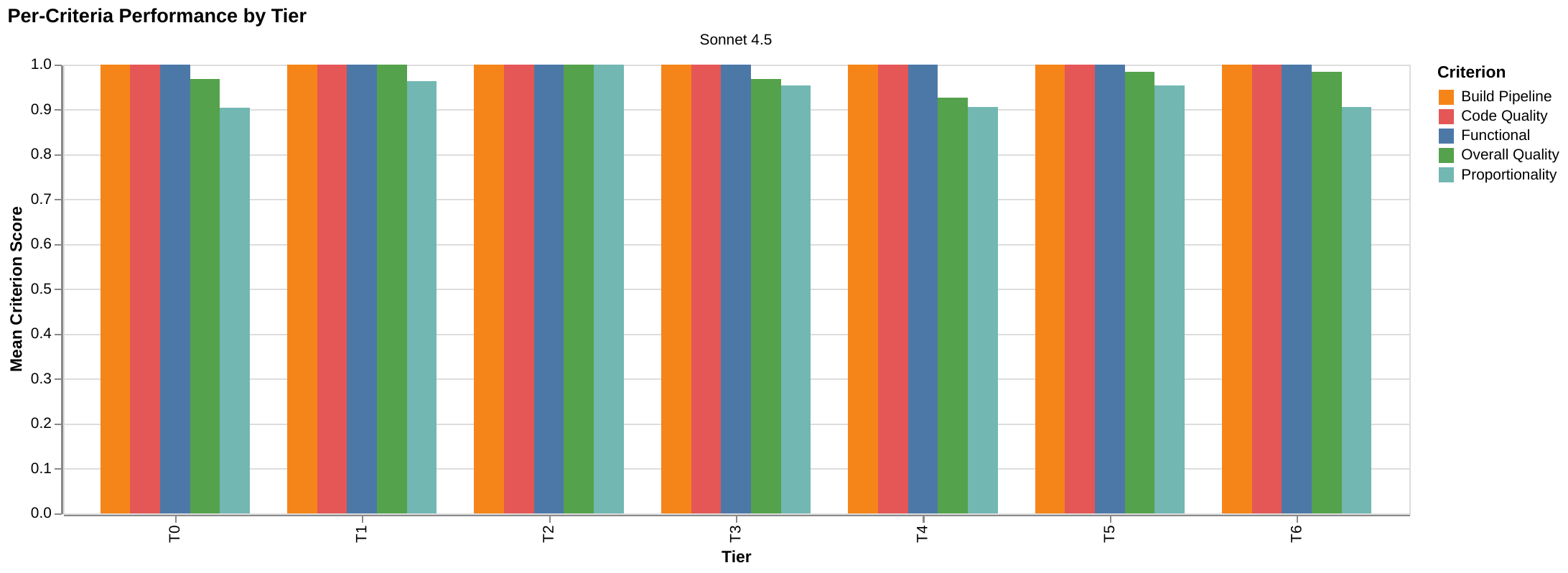}
\caption{Per-criteria scores by tier. Grouped bar chart showing mean scores for five weighted categories (functional correctness, code quality, proportionality, build pipeline, overall quality) across T0--T6. Perfect functional scores (1.00) contrast with variance in proportionality (0.90--1.00) and overall quality (0.93--1.00).}
\label{fig:fig09_criteria_by_tier}
\end{figure}

Judges score five weighted categories: functional correctness (35\%),
code quality (20\%), proportionality (15\%), build pipeline (10\%), overall quality
(20\%). Table~\ref{tab:tab04_criteria_performance} shows detailed per-criteria performance across tiers, and Figure~\ref{fig:fig09_criteria_by_tier} visualizes the breakdown.

\begin{table}[htbp]
\centering
\caption{Per-Criteria Performance Comparison}
\label{tab:tab04_criteria_performance}
\begin{tabular}{lrl}
\toprule
Criterion & Weight & Sonnet 4.5 Mean ($\pm\sigma$) \\
\midrule
functional & 0.35 & 1.000 $\pm$ 0.000 \\
code\_quality & 0.20 & 1.000 $\pm$ 0.000 \\
proportionality & 0.15 & 0.940 $\pm$ 0.083 \\
build\_pipeline & 0.10 & 1.000 $\pm$ 0.000 \\
overall\_quality & 0.20 & 0.975 $\pm$ 0.039 \\
\bottomrule
\end{tabular}
\end{table}

All tiers score 1.00 on functional criteria (file exists, correct output,
exit code 0). Perfect scores suggest the task is trivially easy.

The largest score differences appear in subjective categories. Proportionality:
T6 scored lower because judges noted cache artifacts (\texttt{.ruff\_cache},
\texttt{.pytest\_cache}) remaining in workspace. Overall quality: subjective engineering
judgment shows the most variance across judges.

Build pipeline: all tiers score 1.00, indicating clean execution.

Having presented the quantitative results from the dryrun validation, I now turn to interpretation and implications for future evaluation at scale.

\section{Discussion}\label{sec:discussion}

The dryrun is not very useful for serious analysis, but I will dive into
what I learned about the framework's behavior on this trivially simple task,
while being honest about the limitations inherent in N=1 experiments and ceiling
effects.

\subsection{What the dryrun Tells Us}

The Hello World task is, by design, trivially easy. All seven tiers score grade
A with mean scores between 0.943-0.983. This validates exactly what I said in
Section~\ref{sec:methodology}: "Even T0 should nail this test." And it did.

\textbf{Ceiling effect dominates}: When quality converges at near-perfect levels, we
cannot differentiate tiers by capability. T0's empty prompt (subtest 00 uses no
system prompt at all) and T6's maximal configuration (61 skills + all tools + 44
agents) produce equivalent functional output. This is exactly what we expect for
Hello World, no amount of architectural sophistication helps when the task
requires a single \texttt{print()} statement.

\textbf{Cost differentiation still works}: Despite quality convergence, Cost-of-Pass
varies 3.8$\times$ from \$0.065 (T5) to \$0.247 (T6). This demonstrates the framework's
ability to measure economic trade-offs even when quality metrics saturate. On
more complex tasks with quality variance, both dimensions differentiate.

\textbf{Pipeline validation successful}: The framework executed all seven tiers,
collected 21 judge evaluations, computed consensus scores, generated 24 figures
and 10 tables, and produced structured CSV exports. All components worked as
designed.

\subsection{Cost-Performance Trade-offs}\label{sec:cost-tradeoffs}

The dryrun reveals hints of a pattern: more is not always better.

T5 achieves Frontier CoP through selective feature loading, it combines T1's
efficient skills with T3's delegation patterns but avoids T6's "everything
enabled" overhead. T5's cache creation tokens (4,629) are 5-10x lower than other
tiers (23,106-44,337), directly explaining its cost advantage.

T6 costs the most (\$0.247) despite scoring the lowest
(0.943). Loading 61 skills + all tools + 44 agents actually made things worse.
Judges explicitly noted cache artifacts and unnecessary complexity. This lines
up with the hypothesis that prompt complexity hurts quality when the task is in
the model's training set.

T4's hierarchical overhead is another example. T4 costs 30\% more than T3
(\$0.168 vs \$0.129) for this trivial task. The self-correction loops and nested
orchestration add latency (41.2s vs 29.9s) without improving quality. On complex
tasks needing iterative refinement, maybe T4 justifies the overhead. On simple
tasks, it is pure waste.

See Section~\ref{sec:token-analysis} for detailed token analysis. T2 (tooling) shows schema loading bloat with 137K total tokens versus T1's 115K. Skills-based approaches (T1, T3) stay lean while still
enabling domain knowledge.

Bottom line for production: match tier complexity to task complexity. Do not use
T6 for trivial tasks. Do not use T0 for tasks needing specialized tools or
multi-step reasoning. T5's hybrid approach seems to be optimal, load features
selectively based on what the task actually needs, do not just maximize
everything.

\subsection{Judge Behavior}\label{sec:judge-behavior}

The 3-judge consensus mechanism reveals interesting patterns.

Haiku awards S grades more frequently, 5 out of 7 tiers got perfect scores. Scores
range 0.93-1.00, and Haiku consistently scores higher than Opus or Sonnet.

Opus never awards S grades. Scores range 0.93-0.96, consistently the toughest
judge. Opus reliably deducts points for cache artifacts that Haiku overlooks.

Sonnet splits the difference. Awards S grades in 4/7 tiers (T2, T3, T4, T5),
scores range 0.90-1.00.

Given that the results in most cases are a single line, the 1.0 grade is
incorrect and points to agents being a little too lenient. Maybe some prompt
tweaks will fix this, but that also can be due to the simplicity of this task.
This can be investigated in future analysis.

Inter-rater agreement is predictably low: Krippendorff's $\alpha$ = -0.117. But that is
expected with N=1 and near-perfect scores. On tasks with more variance,
agreement should improve as judges separate clear failures from clear successes.

Despite the disagreement, the 3-judge mean works. When Haiku awards 1.00 and
Opus awards 0.93, the mean captures the true quality without getting pulled to
either extreme. This validates the multi-judge consensus design.

One scaling problem: judge time dominates total latency. 77-86\% of execution
time is judge evaluation (128-178s), not agent execution (25-41s). With 3 judges
per run, judge costs are 3x per evaluation. For large-scale experiments (N=10 $\times$
113 subtests = 1,130 runs $\times$ 3 judges = 3,390 judge evaluations), judge cost uses
the budget fast. Future work will explore single-judge evaluation,
confidence-based selection (use Opus only when Sonnet/Haiku disagree), evaluate
if prompt improvements can get the cheaper Haiku model to be an effective judge,
or give different prompts to the same judge model.

\subsection{Limitations}

\textbf{N=1 prevents inferential statistics}. With only one run per tier, I cannot
compute standard deviations, confidence intervals, or significance tests. All
tier comparisons are point estimates. A single outlier run could flip all
conclusions. The analysis pipeline generated 24 figures and 10 tables, correctly reports \texttt{nan} for standard
deviation and sets confidence intervals to (point, point). Statistical warnings
appear in the output to make clear that results are not expected to be robust.
This is a limitation of this run, not the framework itself.

\textbf{Single task, trivial complexity}. Hello World does not need skills, tools,
multi-agent coordination, or hierarchical reasoning. The dryrun validates the
pipeline works, not whether architectural complexity improves quality on hard
tasks.

\textbf{Single model}. All agent runs use Sonnet 4.5. I have not tested whether tier
rankings hold for Opus 4.5, Haiku 4.5, or other model families.

\textbf{No thinking mode variants}. The dryrun uses standard inference without extended
thinking. Models with thinking enabled might show different cost-quality
trade-offs.

\textbf{Ceiling effect masks capability differences}. When all tiers score 0.94-0.98, I
cannot tell which architecture would excel on harder tasks. The full experiment
(113 subtests including complex multi-file repos) will differentiate
capabilities.

\textbf{Judge evaluation time bottleneck}. 3 sequential judges per run creates a 3x cost
multiplier. Parallel judge execution would reduce latency but not cost.

\section{Conclusions}

This paper introduced the Scylla framework, and shows that it works, end-to-end.
All seven tiers executed successfully, three judges scored everything, and the
analysis pipeline produced figures and tables automatically. The dryrun
validates the methodology on the simplest possible task, Hello World, before I
scale up to complex multi-file repos. What is missing is review and feedback
from others, which is what this paper helps enable.

What did I learn? Five things stand out:

\begin{enumerate}
\item The framework is operational.
\item Quality converges on trivial tasks, making the framework overkill.
\item All tiers scored grade A, suggesting that throwing more complexity at Hello World does not help, which was expected. That obviousness is what makes it a good pipe-cleaner run.
\item Cost still varies 3.8x despite identical quality, showing the framework can measure economic trade-offs even when quality saturates. T5's hybrid approach achieves Frontier CoP by selectively loading features instead of maximizing everything.
\item And the Token Efficiency Chasm I hypothesized in Section~\ref{sec:tiered-ablation}? The data is consistent with this, as T6 burns nearly double the cache read tokens (218K vs 113K) compared to T0.
\end{enumerate}

Did I answer my original questions? Partially. CoP lets me quantify efficiency;
T5 is substantially cheaper than T6 despite equivalent quality. On this task, the sum is
\emph{not} more than the parts; T6 scores lowest despite highest cost. But the hard
questions need harder tasks, I cannot tell if any tier dominates universally from
a single Hello World run, and I have not tested model-to-model comparisons yet.
That work is left for a future exercise.

What about my hypotheses? The KISS principle hypothesis has preliminary support, maximal complexity (T6) scores worst on this training-set-likely
task. But I have not tested inverse KISS on out-of-distribution tasks yet, and
specialization advantages (H1) are inconclusive because Hello World does not
require delegation or tools.

There is no real practical takeaway yet, since the testing was insufficient to
come to any real conclusions. Answering those questions is left for the next
exercise, and this framework can be used for doing so.

\section{Further Work}\label{sec:further}

The dryrun validates the framework works. Now it is time to scale up and fill in
the gaps.

\textbf{Full-scale experiments}: Run the complete test001 dataset with (N=10, 113
 subtests, 1,130 runs total). Running the analysis will start to enable valid
statistical inference about the relationship between prompts and the tools.

\textbf{Task diversity}: The dryrun only covers Hello World. The full test suite
includes 46 additional tasks across greenfield (Flask APIs, CLI tools),
brownfield (feature additions to existing repos), refactoring (extract function,
eliminate duplication), bug fixes (off-by-one errors, race conditions), and
documentation (README generation). Running these will show whether tier rankings
hold across workflow categories or if certain tiers excel at specific task
types.

\textbf{Cross-vendor and cross-model evaluation}: The framework is model-agnostic by
design. I would love to extend support to other tools, but right now just doing
analysis on Claude Code alone is hitting my budgets for experimentation
extremely quickly. Setting up local models and accessing tools using these
models will allow more experimentation, but I do not have access to that kind of
compute within my budget at the moment.

\textbf{Advanced analysis}: I am by no means a statistician, and choices I have made
here might be incorrect. My current analysis uses frequentist statistics. There
are more advanced analyses that I am learning about that could help analyze the
flood of data more efficiently. There are also other metrics and data points that
could be useful in this analysis that I am not collecting. I also can save the
runs and do longitudinal studies to see if the results change consistently over
time.

Given the scale and scope of this task, it is going to be an ongoing effort of
learning, testing, and analyzing.

\section*{Acknowledgements}

This work was self-funded by the author. Special thanks to Tuan Nguyen for
reviewing early drafts of this paper and providing valuable feedback.

\bibliographystyle{plain}
\bibliography{references}

\appendix

\section{Detailed Metric Definitions}

Core quality and economic metrics are defined in Section~\ref{sec:metrics}. This appendix provides additional metrics used for future analysis. For complete definitions including instrumentation details, see the repository at \url{https://github.com/HomericIntelligence/ProjectScylla/blob/main/.claude/shared/metrics-definitions.md}.

\textbf{Change Fail Percentage (CFP)}: Proportion of code changes that cause failures.

Formula: $\text{CFP} = \frac{\text{failed\_changes}}{\text{total\_changes}}$

Range: [0, 1], lower is better. Measures production stability.

\subsection{Process Metrics}

\textbf{Latency}: Time from query submission to response completion, measured in seconds. Components include Time-to-First-Token (TTFT), total response time, and tool execution time (if applicable).

\textbf{Strategic Drift}: Deviation from original goal over multi-step tasks.

Measurement: $\text{Strategic\_Drift} = \text{cosine\_distance}(\text{initial\_goal\_embedding}, \text{final\_action\_embedding})$

Range: [0, 2], where 0 = perfect goal alignment, 2 = completely opposite direction.

\textbf{Ablation Score}: Isolated contribution of a single component to overall performance.

Formula: $\text{Ablation\_Score} = \text{performance\_with\_component} - \text{performance\_without\_component}$

Positive values indicate the component improves performance, negative values indicate harm, near-zero indicates no effect.

\subsection{Statistical Reporting}

Always report metrics with: (1) point estimate, (2) confidence interval (95\% CI recommended), (3) sample size (n), and (4) comparison p-value (if comparing tiers).

Example: Pass-Rate: 0.67 (95\% CI: 0.54--0.80), n=50

\section{Data Dictionary and Generated Outputs}

All raw data, figures, and tables are available in the repository at \texttt{docs/arxiv/dryrun/}. The dataset includes 7 runs (one per tier), 21 judge evaluations (3 judges per run), and 105 criteria scores (5 criteria per judge), organized across 24 figures and 10 tables. Data files include \texttt{runs.csv}, \texttt{judges.csv}, \texttt{criteria.csv}, and \texttt{summary.json}. All figures are available in PNG/PDF/Vega-Lite/CSV formats for reproduction and analysis.

\textbf{Repository}: \url{https://github.com/HomericIntelligence/ProjectScylla}

\section{Reproducibility Checklist}

\textbf{Repository}: \url{https://github.com/HomericIntelligence/ProjectScylla}

\textbf{Key Configuration Files}:

\begin{itemize}
\item Tier definitions: \texttt{config/tiers/tiers.yaml}
\item Opus Model: \texttt{config/models/claude-opus-4-5.yaml}
\item Sonnet Model: \texttt{config/models/claude-sonnet-4-5.yaml}
\item Haiku Model: \texttt{config/models/claude-haiku-4-5.yaml}
\item Judge system prompt: \texttt{config/judge/system\_prompt.md}
\item Test definitions: \texttt{tests/fixtures/tests/*/test.yaml}
\item Rubric schemas: \texttt{tests/fixtures/tests/*/expected/rubric.yaml}
\end{itemize}

\textbf{Required Software}:

\begin{itemize}
\item Pixi (package manager)
\item Claude Code CLI (requires Anthropic API key or Claude Max subscription)
\item Python 3.12+
\end{itemize}

\textbf{Execution Steps}:

\begin{lstlisting}
# 1. Clone repository
git clone https://github.com/HomericIntelligence/ProjectScylla
cd ProjectScylla

# 2. Install dependencies
pixi install

# 3. Run evaluation (example for test-001, tier T0)
pixi run python scripts/run_e2e_experiment.py \
  --test tests/fixtures/tests/test-001 \
  --tier T0 \
  --runs 10

# 4. Generate figures and tables
pixi run python scripts/generate_figures.py \
  --results <output_directory>
pixi run python scripts/generate_tables.py \
  --results <output_directory>
\end{lstlisting}

\end{document}